\documentclass[12pt, a4paper]{article}

\usepackage[latin1]{inputenc}

\usepackage{pdfpages}
\usepackage{bm}
\usepackage{upgreek}
\usepackage{natbib}
\usepackage{rotating}

\usepackage{verbatim}
\usepackage[margin=3cm]{geometry}
\usepackage{url}
\usepackage{fancyhdr}
\usepackage{color}
\usepackage{amsmath, amssymb}
\usepackage{longtable}
\usepackage{lscape}

\newcommand{\cpo}{\text{CPO}}

\newcommand{\qedbox}{\hfill\vcenter{\hrule height.5pt\hbox{\vrule
width.5pt height8pt \kern8pt\vrule width.5pt}\hrule height.5pt}}

\newcommand{\given}{\,\vert\,} % for "X gegeben Y" also $X\given Y$ schreiben

\title{\bf Sensitivity analysis for Bayesian hierarchical models}

\author{{Ma{\l}gorzata Roos$^{\text {a}}$, Thiago G. Martins$^{\text {b}}$, Leonhard Held$^{\text {a}}$ $\&$ H{\aa}vard Rue$^{\text {b}}$}\\
{\small $^{\text {a}}$ Division of Biostatistics, Institute for Social and Preventive Medicine,}\\
{\small University of Zurich, Hirschengraben 84, CH-8001 Zurich, Switzerland}\\
{\small $^{\text {b}}$ Department of Mathematical Sciences, Norwegian University for Science and Technology,}\\
{\small N-7491 Trondheim, Norway}\\
}

\date{\today}

\begin{document}
\maketitle

% =======================================
%\section{Abstract} \label{Abstract}
% =======================================

\noindent{\bf Abstract}
Prior sensitivity examination plays an important role in applied Bayesian analyses.
This is especially true for Bayesian hierarchical models, where interpretability of the parameters within deeper layers in the hierarchy becomes challenging.
In addition, lack of information together with identifiability issues may imply that the prior distributions for such models have an undesired influence on the posterior inference.
Despite its relevance, informal approaches to prior sensitivity analysis are currently used.
They require repetitive re-runs of the model with ad-hoc modified base prior parameter values.
Other formal approaches to prior sensitivity analysis suffer from a lack of popularity in practice, mainly due to their high computational cost and absence of software implementation.
We propose a novel formal approach to prior sensitivity analysis which is fast and accurate.
It quantifies sensitivity without the need for a model re-run.
We develope a ready-to-use {\tt priorSens} package in {\tt R} for routine prior sensitivity investigation by {\tt R-INLA}.
Throughout a series of examples we show how our approach can be used to detect high prior sensitivities of some parameters as well as identifiability issues in possibly over-parametrized Bayesian hierarchical models.


%{\textcolor{red}{For computations presented here R version 3.0.1 with 64-bit together with INLA (Mon 22 Jul 2013) and priorSens package downloaded from the Dropbox on 20131204 was used.}}
%{\textcolor{red}{Parameters were set to {\tt grid$\_$epsilon} $=$ 0.00354 for the value of the Hellinger distance of the grid around the base prior values, which corresponds to 0.01 unit variance normal mean, and to {\tt total$\_$n$\_$out} $=$ 400 for the number of points on the grid.}}

% =======================================
\section{Introduction} \label{S1}
% =======================================

Nowadays applied statisticians have a wealth of both frequentist and Bayesian procedures at their disposal.
The prominent feature of the latter approach is its ability to incorporate prior knowledge in the analysis.
This feature, however, is both a benefit and a challenge.
A Bayesian model is said to be sensitive (non-robust) with respect to the prior distribution if its posterior distribution dramatically changes when the base prior parameter values are altered slightly.
Recently, implementation of the hierarchical framework has lead to the development of increasingly intricate models.
Unfortunately, their complexity makes an extensive elicitation of the base prior for each hierarchy layer practically impossible.
Instead, base priors tend to be determined in a rather casual fashion and without appropriate reflection leading to arbitrary and inaccurately specified parameter values.
At the same time, due to model complexity, the impact of possibly misspecified prior parameter values on outputs is unclear.
In addition, the adequacy of the sample size needed for a reliable estimation of each layer is, in fact, unknown.
It can happen that models are ovarparametrized \citep{carlin_louis_98} and it may not be obvious to decide which parameters are well identified by the data and which are not \citep{dawid_79, gelfand_sahu_99, eberly_carlin_00}.
Hence, development of complex Bayesian models without any prior robustness diagnostics may be problematic.
In order to ensure reliable (robust) results, it is crucial to verify how sensitive the resulting posteriors are for each prior input.

The relevance of sensitivity and uncertainty analyses for exploring complex models has been highly emphasized in the literature (\citet{saltelli_chan_scott_00}, \citet{cacuci_03}, \citet{oakley_ohagan_04}, \citet{saltelli_tarantola_campolongo_ratto_04}, \citet{cacuci_ionescubujor_navon_05} and \citet{saltelli_ratto_andres_campolongo_cariboni_gatelli_saisana_tarantola_08}).
In Bayesian statistics the general sensitivity concept involves broad issues like influential observations, uncertainty of the sampling model and prior inadequacy (\citet{geisser_92}, \citet{lavine_92}, \citet{geisser_93}, \citet{gustafson_wasserman_95}, \citet{clarke_gustafson_98}, \citet{millar_stewart_07}, \citet{zhu_ibrahim_tang_11} and \citet{ibrahim_zhu_tang_11}).

\subsection{Bayesian formal sensitivity analysis}

Sensitivity to the prior parameter specifications is a crucial part of the general sensitivity setting \citep{berger_riosinsua_ruggeri_00, riosinsua_ruggeri_martin_00, ruggeri_08}, as inadequate prior parameter specifications can lead to distorted findings for both influential observations and uncertainty of the sampling model.
To date two approaches to sensitivity analysis can be distinguished: the global and the local one.
The global approach considers the class of all priors compatible with the elicited prior information and computes the range of the posteriors as the prior varies over the class.
This range is typically found by determining the ``extremal" priors in the class that yield maximally distant posteriors, without explicitly carrying out the analysis for every prior in the class.
In contrast, the local sensitivity approach is interested in the rate of change in posterior with respect to changes in the prior, and usually uses differential calculus to approximate it.
Despite its desirability the global approach is impractical in the Bayesian hierarchical framework whereas the local one is the method of choice \citep{gustafson_00, sivaganesan_00, zhu_ibrahim_lee_zhang_07, perez_martin_rufo_06, mueller_12}.

Local sensitivity approach routinely applied in complex Bayesian hierarchical models can spot which model components are hard to learn from the data and makes the researcher aware of which prior to focus on at the stage of the model construction.
It can be employed for a quick identification of priors that may need more careful attention.
Indeed, there is a strong need for investigation not only of the local worst-case sensitivity but also of the circular sensitivity around a particular base prior values specification, since the analyst might be interested in specific directions on the hyperparameter space \citep{kadane_for_geisser_92}.

For local Bayesian robustness investigations a variety of frameworks can be distinguished \citep{gustafson_00}.
They differ according to which posterior results are used (distribution or summaries), what kinds of prior perturbations are used (geometric or parametric), whether the worst-case sensitivity is measured in the absolute or relative sense and what classes of discrepancy measures are considered.
In particular, \citet{mcculloch_89} following \citet{cook_86} approximated prior worst-case robustness by the principal eigenvalue of an appropriate infinitesimal ratio.
This approach has been further refined by \citet{zhu_ibrahim_lee_zhang_07}, \citet{zhu_ibrahim_tang_11}, \citet{ibrahim_zhu_tang_11} and \citet{mueller_12}.
Alternatively, \citet{weiss_cook_92} suggested a graphical approach for assessing posterior influence.
Other advances to local Bayesian robustness can be found in \citet{kass_tierney_kadane_89}, \citet{weiss_96}, \citet{dey_birmiwal_94}, \citet{vanderLinde_07} and \citet{roos_held_11}.

\subsection{Informal approaches and dedicated software}

Surprisingly, despite considerable theoretical contributions to formal sensitivity analysis, their every-day application is not guaranteed at all.
Nowadays, in the few cases when the lack of prior robustness is assessed, brute force and informal approaches are used instead.
An informal technique consists of repetitive runs of the model with ad hoc modified prior inputs.
If the posteriors subjectively do not differ much, non-sensitivity (robustness) is claimed.
The main drawback of such an approach is that it requires several re-runs of the model, which may be extremely time consuming.
The informal approach lacks instructions how the input modifications should be performed and how the differences in the results should be judged in a standardized way.
Consequently, in order to guarantee reproducibility of Bayesian robustness considerations, the use of a formal sensitivity approach is highly desirable.

Although the need and importance of a formal prior robustness investigation have been ubiquitously approved, its popularity deficit in practice seems mainly due to the non-existence of such a facility in current Bayesian programs \citep{ruggeri_08}.
Hence, a development of a formal robustness methodology, which is feasible, fairly quick, operating with low extra computing effort and provided by default in a dedicated software, is strongly required \citep{berger_riosinsua_ruggeri_00, lesaffre_lawson_12}.
Furthermore, in order to become widely used, its compatibility with the MCMC \citep{gilks_richardson_spiegelhalter_96} framework is welcome.

\subsection{Scope of paper}

In this paper we suggest the use of a Bayesian formal local circular sensitivity, which can be conveniently applied to Bayesian hierarchical models.
The novelty of our approach hinges on the choice of a grid, a set of base prior parameter specifications modified in a standardized way.
Our approach guarantees a nearly instantaneous sensitivity assessment without any need for a model re-run.
Because our local sensitivity approach operates with low extra computing effort, it is a convenient measure for an every-day use.
In fact, we have created a {\tt priorSens} package in {\tt R} facilitating the use of sensitivity measure described here for routine application in {\tt inla} \citep{rue_martino_chopin_09}.

The remainder of this article is organized as follows: Section~2 defines the sensitivity measure, its calibration with respect to the normal distribution with unit variance and discusses its general implementation and a particular one within {\tt inla}.
Although our local robustness approach is generally applicable its performance for a range of applications with increasing complexity and several latent models is presented in Section~3. 
In these examples we show how to use the proposed methodology in practice to identify sensitive parameters.
Some concluding remarks are given in Section~4.
Two appendices in Sections~5 and 6 provide a proof and review {\tt R-INLA} framework.
Additional findings are reported in Supplementary Material.

% =======================================
\section{Local sensitivity} \label{S2}
% =======================================

% =======================================
\subsection{Definition} \label{S2.1}
% =======================================

We define the local circular sensitivity $S^{c}_{\bm{\gamma}_0}(\epsilon)$ as the set of ratios
\begin{equation} \label{eq1}
S^{c}_{\bm{\gamma}_0}(\epsilon)=\Bigl\{{\text{d}(\pi_{\bm{\gamma}}(\theta|\bm{y}),\pi_{\bm{\gamma}_0}(\theta|\bm{y}))\over \epsilon},\ \text{for}\ \bm{\gamma}\in\text{G}_{\bm{\gamma}_0}(\epsilon)\Bigr\},
\end{equation}
with the grid (contour line) $\text{G}_{\bm{\gamma}_0}(\epsilon)$ of parameter values specifications defined by
\begin{equation} \label{eq2}
\text{G}_{\bm{\gamma}_0}(\epsilon)=\{\bm{\gamma}: \text{d}(\pi_{\bm{\gamma}}(\theta),\pi_{\bm{\gamma}_0}(\theta))=\epsilon\},
\end{equation}
where $\text{d}(\cdot, \cdot)$ denotes a convenient discrepancy measure between two densities.
In our definition the distributional assumption of the prior $\pi_{\bm{\gamma}}(\theta)$ for one particular component $\theta$ of the Bayesian hierarchical model is held fixed and only its parameter values $\bm{\gamma}$ are allowed to vary.
In particular, we denote by $\pi_{\bm{\gamma}_0}(\theta)$ and $\pi_{\bm{\gamma}_0}(\theta|\bm{y})$ the base prior density with parameter values fixed at $\bm{\gamma}_0$ and the resulting marginal posterior density for $\theta$, respectively.

In practice we use a fixed small $\epsilon$ for sensitivity evaluation instead of its infinitesimal approximation.
We suggest detailed exploration of local geometry implied by $\text{d}(\cdot, \cdot)$ in the space of prior distributions and a numerical search for a prior parameter value grid $\text{G}_{\bm{\gamma}_0}(\epsilon)$ with center set at $\bm{\gamma}_0$ and the distance value kept fixed to $\epsilon$ according to Equation~(\ref{eq2}).
Our circular approach naturally adjusts for possible non-orthogonalities of the prior parametrisation, as it examines all directions in the space of prior parameter values on equal footing.

Circular sensitivity can be conveniently summarized by a single number.
For example, the worst-case sensitivity $S_{\bm{\gamma}_0}(\epsilon)$ is defined to be the maximum of the circular sensitivity $S^{c}_{\bm{\gamma}_0}(\epsilon)$
\begin{equation} \label{eq3}
S_{\bm{\gamma}_0}(\epsilon)=\text{max}\bigl\{S^{c}_{\bm{\gamma}_0}(\epsilon)\bigl\}=\max\limits_{\bm{\gamma}\in\text{G}_{\bm{\gamma}_0}(\epsilon)}{\text{d}(\pi_{\bm{\gamma}}(\theta|\bm{y}),\pi_{\bm{\gamma}_0}(\theta|\bm{y}))\over \epsilon}.
\end{equation}
In this paper we mainly concentrate on the worst-case sensitivity even though several alternative estimates such as mean, median or minimum could be also reported.

For complex Bayesian hierarchical models sensitivity of each model component at the base prior parameter specification is assessed separately according to Equations~(\ref{eq1})--(\ref{eq3}).
The only input required for sensitivity estimation is the marginal posterior density estimate and the base prior distribution specification.
In the first step, the worst-case robustness $S_{\bm{\gamma}_0}(\epsilon)$ can be checked.
Its high value indicates that a particular prior has to be investigated with more care.
Possible reasons might be a displaced prior caused by misspecified prior parameter values or misspecified prior distribution leading to a prior-data conflict \citep{box_80, evans_moshonov_06} and insufficient sample size at the hierarchy level under consideration.
At this step any other circular sensitivity summary such as mean, median or minimum can be also taken into consideration.

In the second step, the circular sensitivity $S^{c}_{\bm{\gamma}_0}(\epsilon)$ in all directions around $\bm{\gamma}_0$ can be referred to.
Circular sensitivity can be easily depicted, as will be shown later.
The circular sensitivity plots indicate directions in which the most pronounced sensitivity value modification was found.
Their shape depends on the prior distribution, the base prior parameter specification and the model assumed.
The choice of the base prior values specification at the stage of the model construction can be conveniently guided by circular sensitivity and its summary values.

As our approach resorts to posterior and prior densities directly, some sort of a convenient discrepancy measure $\text{d}(\cdot, \cdot)$ to quantify the discrepancy between two distributions is required \citep{gustafson_00}.
One possible choice could be $\phi$-divergence (called also $f$-divergence) between two densities $\pi_{0}$ and $\pi_{1}$ defined as
$$\text{D}_{\phi}(\pi_{1},\pi_{0})=\int \pi_{0}(u)\phi\Bigl({\pi_{1}(u)\over \pi_{0}(u)}\Bigr)du,$$
where $\phi$ is a smooth convex function \citep{amari_90, amari_nagaoka_00, dey_birmiwal_94}.
For example the Kullback-Leibler divergence and Hellinger distance are particular cases with $\phi_{KL}(x)=x\log(x)$ for the Kullback-Leibler divergence and with $\phi_{H}(x)=(\sqrt{x}-1)^2/2$ for the Hellinger distance \citep{dey_birmiwal_94}, respectively.
\citet{robert_96} found that the Kullback-Leibler divergence and the Hellinger distance can frequently be used indifferently and opted that Hellinger distance is more natural as a true distribution distance.

Our preference for Hellinger distance \citep{LeCam_86} is motivated by convenience.
The Hellinger distance is clearly advantageous given marginal posterior distributions and prior distributions provided numerically and attaining nonzero values only on a finite discrete set of points \citep{roos_held_11}.
It is a symmetric measure of discrepancy between two densities $\pi_{0}$ and $\pi_{1}$:
$$\text{H}(\pi_{1},\pi_{0})=\sqrt{{1\over 2}\int_{-\infty}^{\infty}\Bigl\{\sqrt{\pi_{1}(u)}-\sqrt{\pi_{0}(u)}\Bigr\}^2du}=\sqrt{1-\text{BC}(\pi_{1},\pi_{0})}.$$
Here, the Bhattacharyya coefficient $\text{BC}(\pi_{1},\pi_{0})=\int_{-\infty}^{\infty}\sqrt{\pi_{1}(u)\pi_{0}(u)}du$ measures affinity of both densities \citep{bhattacharyya_43}.
Note that the Hellinger distance is invariant to any one-to-one transformation (for example logarithmic, inverse or square-root) of both densities \citep{jeffreys_61, roos_held_11}.

We assume throughout the Gaussian distribution to be parametrized by mean $\mu$ and precision $\lambda$ and the gamma distribution with shape $\alpha$ and rate $\beta$ parameters leading to expectation $\alpha/\beta$ and variance $\alpha/\beta^2$.
For both distributions the Hellinger distance between densities with differing parameter values specifications can be computed analytically.

\begin{figure}[h]
\begin{center}
\setkeys{Gin}{height=0.5\textheight, width=0.5\textwidth, keepaspectratio}
\includegraphics{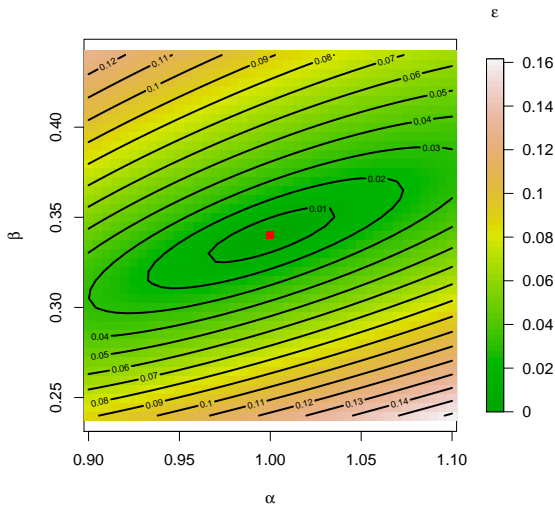}
\end{center}
\caption{Contour plots $\text{G}_{\bm{\gamma}_0}(\epsilon)$ for gamma distribution with center $\bm{\gamma}_0=(\alpha_{0},\beta_{0})=(1,0.34)$ with respect to the Hellinger distance. \label{Ex4geometryGG}}
\end{figure}

\citet{rao_45} discussed the direct correspondence of the Bhattacharyya coefficient and the Fisher information matrix, see also \citet{dawid_77}.
In the context of differential geometry \citet{amari_90} stated that both the Hellinger distance and the Bhattacharyya distance are directly related to the Riemannian distance.
Indeed, the Hellinger distance introduces a non-Euclidean geometry on the space of probability distributions.
As an example consider gamma prior assumed in Section~\ref{S4.4} for the precision of the structured intrinsic conditional autoregressive ``ICAR" (see Section~\ref{S3.4}) component.
Figure~\ref{Ex4geometryGG} shows contour plots $\text{G}_{\bm{\gamma}_0}(\epsilon)$ in Equation~(\ref{eq2}) with respect to the Hellinger distance with center set at $\bm{\gamma}_0^{ICAR}=(\alpha_{0},\beta_{0})=(1,0.34)$.
Equal scaling of x and y-axes highlights that the contours tend to be ellipses rather than circles in Euclidean geometry.

% =======================================
\subsection{Calibration and interpretation} \label{S2.2}
% =======================================

Calibration of differences between two distributions has an advantage that for a particular reference distribution, the experimenter can assess the relevance of the discrepancy in terms of the natural parameter of the benchmark \citep{mcculloch_89, dey_birmiwal_94, goutis_robert_98, roos_held_11}.
Although $\phi$-divergence and Kullback-Leibler divergence have been discussed in the literature, the calibration of the Hellinger distance with respect to the unit variance normal distribution, derived below, seems to be new.
To accomplish this consider $h=h(\mu)=\sqrt{1-\exp(-\mu^2/8)}$, the Hellinger distance between two normal densities with means $\mu_{0}=0$, $\mu_{1}=\mu$ and precisions $\lambda_{0}=\lambda_{1}=1$, and solve it with respect to $\mu$.

\bigskip
\noindent{\bf Lemma:\ }{\it Calibration of the Hellinger distance $h$ between two normal densities N(0,1) and N($\mu(h)$,1), respectively, can be computed as follows:}
$$\mu(h)=\sqrt{-8\log(1-h^2)}$$

Note that $\mu(h)$ is the desired calibration of the Hellinger distance $h$, as $h$ between any two densities is the same as that between N(0,1) and N($\mu(h)$,1).
Given the Hellinger distance $h$ between any two densities, we can quantify discrepancies between them, in terms of the differences in mean from 0 to $\mu(h)$ for normal distribution with unit standard deviation.

Usually it is expected that the sensitivity estimates in Equations~(\ref{eq1}) and (\ref{eq3}) attain values smaller than 1, which indicates that the marginal posteriors obtained for two differing parameter values specifications are closer than both priors assumed.
In such a case data are able to modify priors and lead to closer marginal posteriors.
Sensitivity equal to 1 denotes that marginal posteriors differ the same as the priors, showing that data are unable to modify the prior input at all.
However, in practice both the circular $S^{c}_{\bm{\gamma}_0}(\epsilon)$ and the worst-case $S_{\bm{\gamma}_0}(\epsilon)$ sensitivity estimates can attain values larger than 1.
Possibility of such values has been already attested by \citet{mcculloch_89}, \citet{clarke_gustafson_98}, \citet{plummer_01}, \citet{perez_martin_rufo_06}, \citet{zhu_ibrahim_tang_11} and \citet{mueller_12}.
Whereas \citet{mueller_12} applies truncation at 1 to get rid of excessively large sensitivity values, we prefer to report the unmodified sensitivity estimates.
In case the sensitivity estimate is larger than 1 we will talk about super-sensitivity, using the name coined by \citet{plummer_01}.

In order to get an impression about the relevance of sensitivity values we suggest application of calibration to both the numerator and the denominator of the sensitivity measures defined in Equations~(\ref{eq1}) and (\ref{eq3}).
Interestingly, the ratio of calibrated Hellinger distances in the numerator and denominator can be conveniently approximated by the ratio of Hellinger distances involved in the sensitivity estimates themselves as
$${\mu\bigl(\text{H}(\pi_{\bm{\gamma}}(\theta|\bm{y}),\pi_{\bm{\gamma}_0}(\theta|\bm{y}))\bigr)\over \mu(\epsilon)}\approx {\text{H}(\pi_{\bm{\gamma}}(\theta|\bm{y}),\pi_{\bm{\gamma}_0}(\theta|\bm{y}))\over \epsilon}.$$
Therefore, the sensitivity estimates obtained in applications can be directly interpreted as an approximation of the ratio of calibrated Hellinger distances with respect to the unit variance normal distribution.
Although a particular choice of $\epsilon$ anchors our calibration, the above observation offers an option to interpret the sensitivity magnitude independently of any particular $\epsilon$ value used for the grid computation.
Apart of that, the use of ratio of calibrations leads to its applicability for the whole range of sensitivity values including small and super-sensitivities.

As an example consider the sensitivity values in the last row of Table~\ref{TabEx4Sens} in Section~\ref{S4.4}.
For the ``ICAR" components worst-case sensitivity equal to 0.355 was found.
This value means that the mean change in the unit variance normal distributions for posteriors is about 35.5$\%$ of the mean change of unit variance normal distributions in the prior.
In contrast, super-sensitivity of 1.568 for the unstructured independent random noise ``iid" component shows that the mean change in the unit variance normal distributions for posteriors is about 156.8$\%$ of the mean change in the unit variance normal distributions for priors.

% =======================================
\subsection{Computation} \label{S2.3}
% =======================================

We still have to address two vital topics dealing with the instantaneous computation of posterior density for differing prior parameter values and computation of the sensitivity measure itself.
The general methodology needed for the $\epsilon$-grid search will be explained in the next subsection.

In general, given the marginal posterior density $\pi_{\bm{\gamma}_0}(\theta|\bm{y})$ computed for the base prior $\pi_{\bm{\gamma}_0}(\theta)$, the marginal posterior density $\pi_{\bm{\gamma}}(\theta|\bm{y})$ for the prior $\pi_{\bm{\gamma}}(\theta)$ with a new parameter specification $\bm{\gamma}$ instead of $\bm{\gamma}_{0}$ can be computed instantaneously according to
\begin{equation} \label{instantaneous1}
\pi_{\bm{\gamma}}(\theta|\bm{y})\propto {\pi_{\bm{\gamma}_0}(\theta|\bm{y})\over \pi_{\bm{\gamma}_0}(\theta)}\pi_{\bm{\gamma}}(\theta)
\end{equation}
\citep{tierney_kadane_86, tierney_kass_kadane_89, kass_tierney_kadane_89}.
Formula~(\ref{instantaneous1}) applied to an estimate of the marginal posterior distribution at base prior $\tilde{\pi}_{\bm{\gamma}_0}(\theta|\bm{y})$, provided for example by {\tt inla}, gives
\begin{equation*}
\tilde{\pi}_{\bm{\gamma}}(\theta|\bm{y})\propto {\tilde{\pi}_{\bm{\gamma}_0}(\theta|\bm{y})\over \pi_{\bm{\gamma}_0}(\theta)}\pi_{\bm{\gamma}}(\theta).
\end{equation*}
This general observation makes an instantaneous computation of the Hellinger distance between two marginal posteriors $\tilde{\pi}_{\bm{\gamma}_0}(\theta|\bm{y})$ and $\tilde{\pi}_{\bm{\gamma}}(\theta|\bm{y})$ arising from two slightly shifted prior parameter values $\bm{\gamma}_0$ and $\bm{\gamma}\in\text{G}_{\bm{\gamma}_0}(\epsilon)$ possible as
$$\text{H}(\tilde{\pi}_{\bm{\gamma}}(\theta|\bm{y}),\tilde{\pi}_{\bm{\gamma}_{0}}(\theta|\bm{y}))=\sqrt{1-\text{BC}(\tilde{\pi}_{\bm{\gamma}}(\theta|\bm{y}),\tilde{\pi}_{\bm{\gamma}_{0}}(\theta|\bm{y}))},$$
with
$$\text{BC}(\tilde{\pi}_{\bm{\gamma}}(\theta|\bm{y}),\tilde{\pi}_{\bm{\gamma}_{0}}(\theta|\bm{y}))\approx\int\sqrt{\tilde{\pi}_{\bm{\gamma}}(\theta|\bm{y})\tilde{\pi}_{\bm{\gamma}_{0}}(\theta|\bm{y})}d\theta$$
leading directly to circular sensitivity estimates $S^{c}_{\bm{\gamma}_0}(\epsilon)$ and worst-case sensitivity $S_{\bm{\gamma}_0}(\epsilon)$ in Equations~(\ref{eq1}) and (\ref{eq3}).
We recommend that for marginal posterior densities of precisions the above approach is applied to their log-transformed representations.
The above approach to instantaneous computation of posterior for differing prior parameter specifications makes the necessity of a model re-run superfluous.
Similar computations can be carried out within any framework supporting marginal posterior density $\pi_{\bm{\gamma}_{0}}(\theta|\bm{y})$ estimation.

% =======================================
\subsection{Grid search} \label{S2.4}
% =======================================

For a fixed, small $\epsilon$ the search for the grid $\text{G}_{\bm{\gamma}_0}(\epsilon)$ defined in Equation~(\ref{eq2}) requires exploration of the geometry around the prior parameter values $\bm{\gamma}_0$ in the space of distributions $\pi_{\bm{\gamma}}(\theta)$ (Figure~\ref{Ex4geometryGG}).
The goal is to find the set of prior parameter specifications $\bm{\gamma}$ such that Hellinger distance between $\pi_{\bm{\gamma}}(\theta)$ and the base prior $\pi_{\bm{\gamma}_0}(\theta)$ is equal to $\epsilon$ fulfilling $\text{G}_{\bm{\gamma}_0}(\epsilon)=\{\bm{\gamma}: \text{H}(\pi_{\bm{\gamma}}(\theta),\pi_{\bm{\gamma}_0}(\theta))-\epsilon=0\}$.
In order to find an $\epsilon$ grid for a base prior distribution in, say, two dimensions, a suitable transformation of the Cartesian $(\gamma^{1}, \gamma^{2})$ coordinates to the polar coordinates $(\phi, r)$ is used, where $\phi$ and $r$ denote the angle in radians and modulus, respectively.
For sake of stability of the algorithm $\log(r) = z$ is considered.
We aim for a scaling factor $(\exp(z)\cos(\phi), \exp(z)\sin(\phi))$ which transforms the base prior parameter values $(\gamma_{0}^{1}, \gamma_{0}^{2})$ into an $\epsilon$-distant pair $(\gamma^{1}, \gamma^{2})$ by finding the roots of the analytical equation numerically.

The scaling factor $(\exp(z)\cos(\phi), \exp(z)\sin(\phi))$ obtained is transformed back to Cartesian coordinates using
$$\bigl[\gamma_0^1 + r\cos(\phi)c^x(\phi)\bigr]$$
and
$$\bigl[\gamma_0^2 + r\sin(\phi)c^y(\phi)\bigr],$$
where
\begin{equation*}
c^x(\phi)=\left\{
\begin{array}{rcl}
r^{*}(0) &\ & \mbox{if $\phi \in [-\pi/2, \pi/2]$,} \\
r^{*}(\pi) &\ & \mbox{if $\phi \in [\pi/2, -\pi/2]$}
\end{array}
\right.
\end{equation*}
and
\begin{equation*}
c^y(\phi)=\left\{
\begin{array}{rcl}
r^{*}(\pi/2) &\ & \mbox{if $\phi \in [0, \pi]$,} \\
r^{*}(-\pi/2) &\ & \mbox{if $\phi \in [\pi, 0]$}
\end{array}
\right.
\end{equation*}
with $r^{*}(\delta)$, for radian values $\delta = -\pi/2$, $0$, $\pi/2$, $\pi$, denoting the modulus values obtained at $\delta$ angles during a pre-exploration of the polar coordinate space.
The factors $c^x(\phi)$ and $c^y(\phi)$ are necessary to scale the problem so that $r$ is close to $1$ across different prior distributions.
This practice standardizes the task of computing $\text{G}_{\bm{\gamma}_0}(\epsilon)$, which makes the numerical algorithm more stable and generally applicable.

The polar coordinates approach guarantees that each direction is treated on equal footing as the angles $\phi$ run through an equidistant grid in $[-\pi, \pi]$ interval.
It also implies a natural ordering of the grid points.
This polar approach is applied to both normal and gamma priors used in applications in Section~\ref{S4} but could be easily applied to any other two-parameter prior distribution of interest or even extended to higher dimensions in case the prior has more than two parameters.

% =======================================
\subsection{Local sensitivity in {\tt R-INLA}} \label{S3}
% =======================================

Our fast general circular sensitivity methodology can be implemented without much extra cost by any technique capable of computing marginal posterior distributions, in particular, by the {\tt R-INLA} framework (Appendix~B in Section~\ref{AppendixINLA}).
In practice, however, two settings for the cardinality of the grid $\text{G}_{\bm{\gamma}_0}(\epsilon)$ and the value of $\epsilon$ have to be fixed (Section~\ref{S2.4}).
In applications shown below we consider 400 polar directions and use one particular $\epsilon_0 =$ 0.00354 for the grid search, which corresponds to a unit variance normal distribution with mean equal to 0.01 (Section~\ref{S2.2}).
Table~\ref{TabEx3Eps_d} and the first two examples in Supplementary Material indicate that the exact sensitivity estimates stay stable over a wide range of $\epsilon$ values.
Therefore, there is some room for a tolerable $\epsilon$ choice.

All sensitivity estimates presented below were estimated by the {\tt priorSens} package in {\tt R}.
Our {\tt inla} computations were run for default settings: simplified Laplace strategy for the marginal posterior approximation and central composite design (CCD) for integrating out the hyperparameters.
In general, for precisions log-transformed marginal posterior densities were used.
In addition, for precisions of latent components in {\tt R-INLA} such as ``iid", ``ICAR", the latent Gaussian random walk of the first (``rw1") and the second (``rw2") order an appropriate tuning for their marginal posteriors provided by the function {\tt inla.hyperpar()} was utilized.

% =======================================
\section{Applications} \label{S4}
% =======================================

We start this section by reviewing latent models used in applications.
Next, we demonstrate the use of the circular and worst-case sensitivity for three data sets with increasing hierarchical model complexity.
Additional two examples are discussed in Supplementary Material.

% =======================================
\subsection{Latent models} \label{S3.4}
% =======================================

In the applications provided in following paragraphs we use several latent models in {\tt R-INLA} such as ``iid", ``ICAR", ``rw1", ``rw2" and stochastic partial differential equations (``spde").
Here we describe them shortly.

The unstructured independent random noise model (``iid") for random effects in the vector $\bm{v}$ defines the Gaussian random field to be a vector of independent Gaussian distributed random variables  $v_{j}\stackrel{ind}{\sim} \text{N}(0,\tau_{iid}^{-1})$ with precision $\tau_{iid}$, for which the gamma prior is assumed.
A more involved structured intrinsic conditional autoregressive model (``ICAR" called also ``besag") \citep{besag_york_mollie_91} in component $\bm{u}$ assumes that conditions for neighbouring random effects tend to be similar.
The Gaussian random field $\bm{u}=(u_1, u_2, \ldots, u_n)$ is defined as
$$u_i|u_j, i\neq j, \tau_{ICAR}\sim \text{N}({1\over n_i}\sum_{i\sim j}u_j, {1\over {n_i \tau_{ICAR}}}),$$
where $i\sim j$ indicates that two random effects $i$ and $j$ are neighbours and $n_i$ is the number of neighbouring entities of the $i$th object.
In order to guarantee the identifiability of the intercept the option {\tt constr $=$ TRUE}, a sum-to-zero constraint on each connected component, is used.
For ``ICAR" precision $\tau_{ICAR}$ the gamma prior is assumed.
Both ``iid" and ``ICAR" latent models are used in Sections~\ref{S4.4} and \ref{S4.5}.

The first and second-order intrinsic Gaussian Markov random fields (IGMRFs) are frequently used to model smooth, non-linear functions of covariates in one dimesion \citep{held_rue_2010}.
In both models the precision $\tau$ governs the smoothness of the resulting random effect.
For a detailed description of the ``rw1" latent model see Section~\ref{S4.3}.
\citet[equation (3.39)]{rue_held_05} define the joint density of $\bm{x}|\tau$ for the ``rw2" model by
\begin{equation*}
\pi(\bm{x}|\tau)\propto \tau^{(n-2)/2}\exp\Bigl(-{\tau\over 2}\bm{x}^{T}\mathbf{R}\bm{x}\Bigr),
\end{equation*}
with structure matrix $\mathbf{R}$ determined by the second-order random walk \citep[equation (3.40)]{rue_held_05}.

Recently \citet{lindgren_rue_lindstroem_11} provided an explicit link between continuously indexed Gaussian fields and discretely indexed Gaussian Markov random fields using an approximate stochastic weak solution to linear stochastic partial differential equations (SPDEs).
The approach developed there was successfully applied to global temperature reconstruction and hierarchical spatio-temporal analysis \citep{cameletti_lindgren_simpson_rue_12} and has the potential to model not only Matérn but also non-stationary, non-separable, anisotropic, oscillation, space-time and  multivariate fields on manifolds.
In principle the SPDE analysis consists of five steps: domain triangulation, definition of the SPDE model object, definition of the {\tt inla} function, an inla call and extraction of the results for further use and plotting.
For implementation of the SPDE approach in {\tt inla} together with the step-by-step {\tt R}-code see \url{http://www.r-inla.org/examples/case-studies} and \url{http://www.r-inla.org/examples/tutorials}.
In Section~\ref{S4.5} a ``spde" model makes use of the more detailed information contained in exact point location of observations.
It considers the fine-scale spatial structure of the underlying smooth process \citep{lindgren_13, simpson_lindgren_rue_12a, simpson_lindgren_rue_12b}.

In order to fix the notation we consider covariance function between two individual locations $s_i$, $s_j$ of a spatial field $x(\bm{s})$, of the Mat{\'e}rn form
$$c_{\nu}(s_i, s_j)={\sigma^2\over \Gamma(\nu+d/2)(4\pi)^{d/2}\kappa^{2\nu}2^{\nu-1}}(\kappa||s_i-s_j||)^{\nu} K_{\nu} (\kappa||s_i-s_j||),$$
where $\nu$ is the Mat{\'e}rn smoothness parameter of the random field, $\kappa$ is a scale parameter, $\sigma^2$ is the variance parameter and $K_{\nu}$ is the modified Bessel function of the second kind.
The SPDE approach introduced by \citet{lindgren_rue_lindstroem_11} makes use of the observation that Matérn fields with covariance function defined above are the stationary solutions to the SPDE
$$(\kappa^2-\Delta)^{\alpha\over 2} (\tau x(\bm{s})) = W(\bm{s}),$$
where $\alpha=\nu+d/2$, $\Delta=\sum_{j=1}^{d}{\partial^2\over {\partial s_j^2}}$ is the Laplacian and $W(\bm{s})$ is spatial white noise.
In this model $\alpha$ controls the smoothness of the realisations.
The practical spatial range is governed by $\kappa$ according to the approximation suggested by \citet{lindgren_rue_lindstroem_11}: ${\tt range}\approx\sqrt{8\nu}/\kappa$.
On the other hand the nominal field variance $\sigma^2$ is governed by both parameters $\kappa$ and precision $\tau$ due to
$$\sigma^2={\Gamma(\nu)\over \Gamma(\nu+d/2)(4\pi)^{d/2}\kappa^{2\nu}\tau^2}.$$

In the acute myeloid leukemia survival data set considered in Section~\ref{S4.5} the spatial locations are contained in two dimensions.
Therefore, $d=2$ and $\bm{s}\in\mathbb{R}^2$.
For analysis of the data \citet{lindgren_rue_lindstroem_11} assumed an integer $\alpha=2$ implying that the SPDE only involves second order derivatives.
This implies that the Matérn smoothness parameter $\nu=\alpha-d/2=1$.
In such a case the field variance can be expressed as $\sigma^2=1/4\pi\kappa^2\tau^2$ and ${\tt range}\approx\sqrt{8}/\kappa$.
Therefore, investigation of the sensitivity of marginal posterior distributions of $\tau$ and $\kappa$ to prior values assumptions is of prime importance as they are an inherent part of nominal range and variance estimates.
In order to enhance readability of the results we denote in the sequel both parameters by $\tau_{SPDE}$ and $\kappa_{SPDE}$ respectively.

% =======================================
\subsection{Smoothing time series data} \label{S4.3}
% =======================================

\begin{figure}[h]
\begin{center}
\setkeys{Gin}{height=0.6\textheight, width=0.6\textwidth, keepaspectratio}
\includegraphics{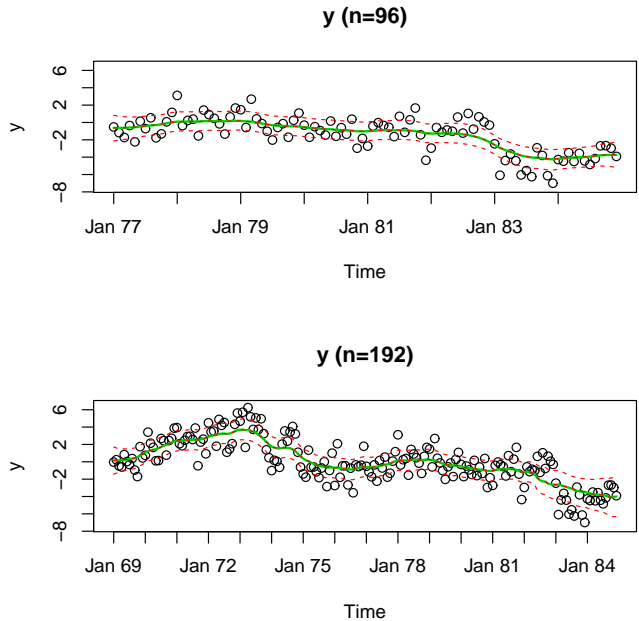}
\end{center}
\caption{Residuals after removal of a seasonal effect in the square-root-transformed monthly deaths and serious injuries counts in road accidents in Great Britain (Jan1969-Dec1984) together with {\tt inla} smoothing according to ``rw1" (posterior mean together with 0.025, 0.5 and 0.975 quantiles): last eight years of the time series $n=96$ (top), full data $n=192$ (bottom). \label{Ex3Data_d}}
\end{figure}

In the first application we consider the time series ($n=192$) on the monthly number of car drivers in Great Britain killed and seriously injured from January 1969 to December 1984 \citep{harvey_durbin_1986, harvey_1989}.
Temporal trends in this time series are of interest, since the seat-belt law became effective on 31 January 1983.
As in \citet{rue_held_05}, we apply the square root transformation to the outcome and consider residuals after removal of the seasonal effect for further analysis.
Counts observed in the last eight years of the time series ($n=96$) are taken to study the influence of the sample size (see Figure~\ref{Ex3Data_d}).

Our main goal is sensitivity estimation of the posterior distribution $\pi( \tau|\bm{y})$ at the base prior parameter specification in a hierarchical model with a latent Gaussian first-order random walk $\bm{x}$ (``rw1") \citep{held_rue_2010}, which can be defined as follows:
Let $\bm{x}|\tau\ \sim \ \text{N}_{n}(\bm{0}, (\tau\mathbf{R})^{-1})$ with the tridiagonal structure matrix $\mathbf{R}$ determined by the first-order random walk \cite[p. 95]{rue_held_05} attaining the following values:
\begin{equation*}
\mathbf{R}_{ij}=\left\{
\begin{array}{rcl}
1 &\ & \mbox{if $i=j=1, n$,} \\
2 &\ & \mbox{if $1<i=j<n$,} \\
-1 &\ & \mbox{if $i=j+1, j-1$,} \\
0 &\ & \mbox{otherwise}
\end{array}
\right.
\end{equation*}
and assume the gamma prior $\text{G}(\alpha,\beta)$ for the hyperparameter $\tau$:
\begin{equation*}
\pi(\tau)={\beta^{\alpha}\over \Gamma(\alpha)}\tau^{\alpha-1}\exp(-\beta\tau).
\end{equation*}
Therefore, by \citet[equation (3.21)]{rue_held_05}
\begin{equation*}
\pi(\bm{x}|\tau)= (2\pi)^{-(n-1)/2}\tau^{(n-1)/2}\exp\Bigl(-{\tau\over 2}\bm{x}^{T}\mathbf{R}\bm{x}\Bigr).
\end{equation*}
Note that rank of $\mathbf{R}$ is $n-1$ and the random walk of first order is an intrinsic GMRF (IGMRF).
Assume that $\bm{y}|\bm{x},\tau\ \sim \ \text{N}_{n}(\bm{x}, (\kappa\mathbf{I})^{-1})$ with $\kappa>0$ fixed.
Therefore,
\begin{equation*}
\pi(\bm{y}|\bm{x}, \tau)= (2\pi)^{-n/2}\kappa^{n/2}\exp\Bigl(-{\kappa\over 2}(\bm{y}-\bm{x})^{T}\mathbf{I}(\bm{y}-\bm{x})\Bigr).
\end{equation*}
It can be shown that
\begin{equation*}
\pi(\tau|\bm{y})= {1\over \text{C}(\alpha,\beta)}\tau^{\alpha+(n-1)/2-1}|\mathbf{Q}|^{-1/2}\exp(-\beta\tau+{1\over 2}\bm{\mu}^{T}\mathbf{Q}\bm{\mu}),
\end{equation*}
where $\mathbf{Q}=\tau\mathbf{R}+\kappa\mathbf{I}$, $\bm{\mu}=\mathbf{Q}^{-1}\kappa\bm{y}^{T}$ and, by an argument provided in in Appendix~A in Section~\ref{Appendix1},
\begin{equation}\label{detRW1}
|\mathbf{Q}|=|\tau\mathbf{R}+\kappa\mathbf{I}|=\prod_{i=1}^{n}(\tau\lambda_{i}+\kappa)=\prod_{i=1}^{n}(\tau(2-2\cos(\pi(i-1)/n))+\kappa).
\end{equation}

\begin{table}
\caption{Worst-case sensitivity estimates for precision $\tau$ as a function of $\epsilon$ for $n=96$ and $n=192$ in Section~\ref{S4.3} at $\bm{\gamma}_{0}=(\alpha_{0},\beta_{0})=(1,0.005)$.}
\centering
\begin{tabular}{lcccc}
\noalign{\medskip}
\hline
\noalign{\smallskip}
$\epsilon$ & $\text{S}_{\bm{\gamma}_0}^{\text{exact}}(\epsilon), n=96$ & $\text{S}_{\bm{\gamma}_0}^{\text{\tt inla}}(\epsilon), n=96$ & $\text{S}_{\bm{\gamma}_0}^{\text{exact}}(\epsilon), n=192$ & $\text{S}_{\bm{\gamma}_0}^{\text{\tt inla}}(\epsilon), n=192$\\
\noalign{\smallskip}
\hline
\noalign{\smallskip}
% latex table generated in R 3.0.1 by xtable 1.7-1 package
% Tue Dec 17 08:01:41 2013
 0.0001 & 0.71 & 0.71 & 0.48 & 0.48 \\
  0.0005 & 0.71 & 0.71 & 0.48 & 0.48 \\
  0.0010 & 0.71 & 0.71 & 0.48 & 0.48 \\
  0.0050 & 0.72 & 0.72 & 0.48 & 0.48 \\
  0.0100 & 0.73 & 0.73 & 0.49 & 0.49 \\ \end{tabular}
\label{TabEx3Eps_d}
\end{table}

\noindent The normalising constant $\text{C}(\alpha,\beta)$ can be computed by numerical integration
\begin{equation*}
\text{C}(\alpha,\beta)= \int_{0}^{\infty}\tau^{\alpha+(n-1)/2-1}|\mathbf{Q}|^{-1/2}\exp(-\beta\tau+{1\over 2}\bm{\mu}^{T}\mathbf{Q}\bm{\mu})d\tau.
\end{equation*}
The equality
\begin{equation*}
\text{H}(\pi(\tau|\bm{y},\alpha_1,\beta_1), \pi(\tau|\bm{y},\alpha_0,\beta_0))=\sqrt{1-{\text{C}((\alpha_0+\alpha_1)/2, (\beta_0+\beta_1)/2)\over \sqrt{\text{C}(\alpha_0,\beta_0)\text{C}(\alpha_1,\beta_1)}}}
\end{equation*}
enables an analytical estimate of the sensitivity.

In order to guarantee the model conjugacy the precision $\kappa$ was fixed at $0.274$, obtained from the residual variance estimate.
We computed both exact and {\tt inla}-driven sensitivity estimates for $\pi( \tau|\bm{y})$ at the base gamma prior with parameter values $\bm{\gamma}_0=(\alpha_0,\beta_0)=(1, 0.005)$.
They agreed perfectly well giving $\text{S}_{\bm{\gamma}_0}^{\text{exact}}(\epsilon_0)=$ 0.48 and $\text{S}_{\bm{\gamma}_0}^{\text{inla}}(\epsilon_0)$ $=$ 0.48 for $\epsilon_0=$ 0.00354, so the mean change in the unit variance normal distributions for posteriors is about 48$\%$ of the mean change in the unit variance normal distributions for priors.
Absolute and relative error ranges of the {\tt inla}-driven circular sensitivity estimates with respect to the analytical ones were equal to (-1.5e-05, 7.1e-05) and (-0.000563, 0.009528), respectively.
Moreover, exact and {\tt inla}-driven estimates were very close for a wide range of $\epsilon$ values and the decreased sample size lead to elevated sensitivity estimates (Table~\ref{TabEx3Eps_d}).

% =======================================
\subsection{Disease mapping} \label{S4.4}
% =======================================

A more challenging non-conjugate and hierarchical example is the analysis of spatial variation of lip cancer in Scotland (1975--1980), which was previously considered by many authors.
Some of the relevant references are \citet{clayton_kaldor_87}, \citet{breslow_clayton_93}, \citet{eberly_carlin_00}, \citet{banerjee_carlin_gelfand_04}, \citet{wakefield_07} and \citet[][Supplementary Material]{fong_rue_wakefield_09}.
Lip cancer in Scotland data set is also used as an example in GeoBUGS User Manual.
Here, we consider observed ($y$) and expected ($e$) cases of lip cancer registered during the time span of six years in each of the $n=56$ counties in Scotland.
We include an intercept $c$ (``const"), a covariable $x$, denoting the proportion of individuals who are employed in agriculture, fishing or forestry scaled by $1/10$, a known offset $\log{e}$ as well as spatial components $\bm{v}$ (``iid") and $\bm{u}$ (``ICAR") described in Section~\ref{S3.4}.

Let $y_{i}$ be realisations of $Y_{i}\given \mu_{i}\stackrel{ind}{\sim} \text{Po}(\mu_{i}),\ i=1,\ldots ,n$ and consider the following six models:
\begin{eqnarray}
\log\mu_{i}&=&\log{e_{i}}+c\qquad\qquad +v_{i}\\
\log\mu_{i}&=&\log{e_{i}}+c\qquad\qquad\qquad +u_{i}\\
\log\mu_{i}&=&\log{e_{i}}+c\qquad\qquad +v_{i} +u_{i}\\
\log\mu_{i}&=&\log{e_{i}}+c+\beta x\quad\ \ +v_{i}\\
\log\mu_{i}&=&\log{e_{i}}+c+\beta x\quad\qquad\ \ +u_{i}\\
\log\mu_{i}&=&\log{e_{i}}+c+\beta x\quad\ \ +v_{i} +u_{i}
\end{eqnarray}

\begin{table}[h]
\caption{Worst-case sensitivity estimates for model components in Section~\ref{S4.4} for $\epsilon_0=$ 0.00354.}
\centering
\begin{tabular}{cccc}
\noalign{\medskip}
\hline
\noalign{\smallskip}
``const" & ``$x$" & ``iid" & ``ICAR" \\
\noalign{\smallskip}
\hline
\noalign{\smallskip}
% latex table generated in R 3.0.1 by xtable 1.7-1 package
% Tue Dec 17 08:01:41 2013
 0.004 &  & 0.244 &  \\
  0.002 &  &  & 0.245 \\
  0.002 &  & 1.587 & 0.274 \\
  0.005 & 0.004 & 0.237 &  \\
  0.004 & 0.004 &  & 0.268 \\
  0.005 & 0.005 & 1.568 & 0.355 \\ \end{tabular}
\label{TabEx4Sens}
\end{table}

\begin{figure}[h]
\begin{center}
\setkeys{Gin}{height=0.7\textheight, width=0.7\textwidth, keepaspectratio}
\includegraphics{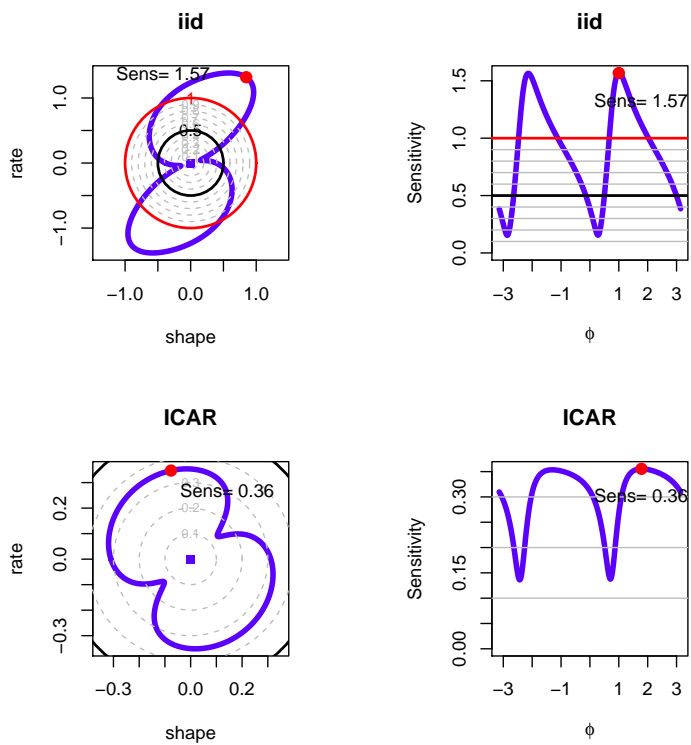}
\end{center}
\caption{Polar sensitivity plots (left) centered at $\bm{\gamma}_{0}$ and sensitivity plots rolled out on the line (right) in each polar direction of the circular $S_{\bm{\gamma}_0}^{c}(\epsilon_0)$ together with the worst-case sensitivity $S_{\bm{\gamma}_0}(\epsilon_0)$ (red dot) for ``iid" (top) and ``ICAR" (bottom) components in model (11) in Section~\ref{S4.4} obtained by {\tt inla} for $\epsilon_0=$ 0.00354. \label{Ex4inla_iid_ICAR}}
\end{figure}

As \citet[][Supplementary Material]{fong_rue_wakefield_09} provided a very careful probabilistic elicitation of the prior values, we adopted their choice here.
For the constant and the regression coefficient of the covariate $\beta$ we assumed normal priors with base prior parameter specification set at $\bm{\gamma}_{0}^{c, \beta}=(0, 0.001)$.
Instead, for the unstructured ``iid" and structured ``ICAR" components we assumed gamma priors for $\tau_{iid}$ and $\tau_{ICAR}$ with base prior parameter values set to $\bm{\gamma}_{0}^{iid}=(1,0.14)$ and $\bm{\gamma}_{0}^{ICAR}=(1,0.34)$, respectively.
Table~\ref{TabEx4Sens} reports the worst-case sensitivities for each component in all six models for the base prior parameter values defined above.
Figure~\ref{Ex4inla_iid_ICAR} shows for precisions of ``iid" and ``ICAR" in model (11) both the polar circular sensitivity plot centered at $\bm{\gamma}_0$ and rolled out on the line with sensitivities (0.1, 0.2,$\ldots$, 0.9, 1) indicated by circles and lines, respectively.
The worst-case sensitivity is marked by a red dot.
Just for better orientation, sensitivity equal 0.5 is pronounced by a black line, whereas the sensitivity value 1 is marked by a red one.

\citet[][Supplementary Material]{fong_rue_wakefield_09} prefer to always include unstructured random effects together with the structured ICAR model, the reason being that since ICAR model contains only a single parameter to govern both the spatial extent of dependence, and the strength of this dependence, there is no place for pure randomness to be accommodated (which can be a problem, particularly if there is negligible spatial dependence).
Their models (0.3), (0.5), (0.4) and (0.6) are complemented with worst-case sensitivity estimates in rows corresponding to Equations~(6), (8), (9) and (11) in Table~\ref{TabEx4Sens}.
We conclude that having both ``iid" and ``ICAR" components at the same time in the model leads for the lip cancer in Scotland data to super-sensitive marginal posteriors for the precision of ``iid" component with respect to the base prior parameter values choice.
Apparently a simultaneous inclusion of both latent components ended in an overparametrized model \citep{carlin_louis_98} and the ``iid" component became non-identifiable \citep{eberly_carlin_00}.

% =======================================
\subsection{Spatial variation in survival data} \label{S4.5}
% =======================================

The analysis of leukaemia data set from \citet{henderson_shimakura_gorst_02} leads to another challenging non-conjugate hierarchical model.
It encodes spatial variation in 1043 cases of acute myeloid leukaemia (AML) survival in adults diagnosed between 1982 and 1998 in north-west England.
Moreover, demographic variables together with the white blood cell count at diagnosis ({\tt wbc}) and the Townsend score measuring deprivation in the district of residence ({\tt tpi}) are recorded there.
Additionally, spatial information in form of 1043 individual point locations of the AML cases together with their affiliation to one of the 24 administrative districts is provided.

We consider the parametric Weibull proportional hazards model with baseline hazard \citep{martino_akerkar_rue_11}
$$h(t|\bm{x})=\alpha t^{\alpha-1}\lambda,$$
where $\alpha$ is the modulus (shape parameter) of the Weibull distribution, a hyperparameter to which a gamma prior distribution is assigned, and the term $\lambda$ is linked to the linear predictor $\bm{x}^{T}\bm{\beta}$ as in one of the following three ways:
\begin{eqnarray}
\log(\lambda)&=&\bm{x}^{T}\bm{\beta} + \bm{v}\\
\log(\lambda)&=&\bm{x}^{T}\bm{\beta}\qquad + \bm{u}\\
\log(\lambda)&=&\bm{x}^{T}\bm{\beta}\qquad \qquad + \bm{w},
\end{eqnarray}
where $\bm{x}$ is a vector comprising the intercept ({\tt const}), {\tt gender}, {\tt age}, {\tt wbc} and {\tt tpi}; $\bm{\beta}$ is the corresponding vector of the fixed effects parameters; $\bm{v}$ and $\bm{u}$ represent the ``iid" and ``ICAR" models for the districts, respectively, and $\bm{w}$ denotes the spatial ``spde" term for each location.

The simplest district-level ``iid" model in $\bm{v}$ component assumes between districts independence.
A more involved district-level ``ICAR" model in component $\bm{u}$ assumes that conditions for AML tend to be similar in neighbouring areas \citep{henderson_shimakura_gorst_02, martino_akerkar_rue_11}.
Political districts are viewed as neighbours if they share a common boundary.
The individual-level ``spde" model for $\bm{w}$ makes use of the more detailed information contained in the individual exact point location of the AML patient's residence.
Therewith it takes the underlying continuous fine-scale spatial risk process into account (see Section~\ref{S3.4} for details).

\begin{table}
\caption{Base prior parameter values for the ``iid", ``ICAR" and ``spde" models considered in Section~\ref{S4.5}.}
\centering
\begin{tabular}{l|l|l|l}
\noalign{\bigskip}
\hline
\hspace{0.001 pt} &  &  & \\
Parameter & ``iid" & ``ICAR" & ``spde"\\
\hspace{0.001 pt} &  &  & \\
\hline
\hspace{0.001 pt} &  &  & \\
$\beta_0, \beta_1,\ldots ,\beta_4$ & N$(0, 0.001^{-1})$ & N$(0, 0.001^{-1})$ & N$(0, 0.001^{-1})$ \\
$\log(\alpha)$ & logG$(0.05, 0.1)$ & logG$(0.05, 0.1)$ & logG$(0.05, 0.1)$ \\
$\log(\tau_{iid})$ & logG(1, 5e-5) &  &  \\
$\log(\tau_{ICAR})$ &  & logG(1, 5e-5) &  \\
$\log(\tau_{SPDE})$ & & & N$(-3.633, 0.1^{-1})$ \\
$\log(\kappa_{SPDE})$ & & & N$(2.368, 0.4^{-1})$ \\
\end{tabular}
\label{TabSPDE}
\end{table}

Our selection of the prior parameter values is based on the choice made by \citet{lindgren_rue_lindstroem_11} (see Table~\ref{TabSPDE}).
Base prior parameter values for intercept $\beta_0$ and the regression coefficients $\beta_1,\ldots ,\beta_4$ as well as for $\tau_{iid}$ and $\tau_{ICAR}$ correspond to the default prior parameter values choice assumed by {\tt inla}.
In addition, we assumed independence of priors for log spatial range ($\log(\kappa_{SPDE})$) and for log precision ($\log(\tau_{SPDE})$).

Table~\ref{TabEx5Sens_and_Sens_smooth_ig_n50nonscaled} (left) shows the {\tt inla}-driven sensitivity estimates for the three models in Equations~(12)-(14) with base prior values specifications given in Table~\ref{TabSPDE}.
The hyperparameters of the ``iid" and ``ICAR" models are super-sensitive.
Marginal posterior distributions of the regression coefficients and for Weibull $\alpha$ for all three models do not show much sensitivity at all with the mean change in the unit variance normal distributions for posteriors being at most 0.7$\%$ of the mean change in the unit variance normal distributions for priors.

\begin{table}
\caption{Worst-case sensitivity estimates for model components in Section~\ref{S4.5} for $\epsilon_0=$ 0.00354. Left: {\tt wbc} and {\tt tpi} considered directly; Right: models ($^{*}$) with {\tt rw1} applied to {\tt wbc} and {\tt rw2} to {\tt tpi} together with discretization in 50 unique equidistant values with default base prior G(1, 5e-5).}
\centering
\begin{tabular}{l|ccc|ccc}
\noalign{\bigskip}
\hline
\rule{0pt}{0pt} &  &  & \\
Parameter & ``iid" & ``ICAR" & ``spde" & ``iid$^{*}$" & ``ICAR$^{*}$" & ``spde$^{*}$" \\
\rule{0pt}{0pt} &  &  & \\
\hline
\rule{0pt}{0pt} &  &  & \\
% latex table generated in R 3.0.1 by xtable 1.7-1 package
% Tue Dec 17 08:01:44 2013
 const & 0.00469 & 0.00452 & 0.00649 & 0.00534 & 0.00520 & 0.00754 \\
  sex & 0.00217 & 0.00216 & 0.00219 & 0.00218 & 0.00218 & 0.00221 \\
  age & 0.00007 & 0.00007 & 0.00007 & 0.00007 & 0.00007 & 0.00007 \\
  wbc | $\log(\tau_{rw1(wbc)})$ & 0.00002 & 0.00001 & 0.00002 & 0.83998 & 0.83007 & 0.82776 \\
  tpi | $\log(\tau_{rw2(tpi)})$ & 0.00031 & 0.00030 & 0.00031 & 0.89031 & 0.87505 & 0.88901 \\
  $\log(\alpha)$ & 0.00715 & 0.00712 & 0.00750 & 0.00721 & 0.00715 & 0.00757 \\
  $\log(\tau_{iid})$ & 1.06591 &  &  & 0.86448 &  &  \\
  $\log(\tau_{ICAR})$ &  & 1.12592 &  &  & 0.94676 &  \\
  $\log(\tau_{SPDE})$ &  &  & 0.18746 &  &  & 0.17783 \\
  $\log(\kappa_{SPDE})$ &  &  & 0.32776 &  &  & 0.33474 \\ \end{tabular}
\label{TabEx5Sens_and_Sens_smooth_ig_n50nonscaled}
\end{table}

Both \citet{kneib_fahrmeir_07} and \citet{martino_akerkar_rue_11} considered more flexible smooth effects of covariates, so we also included {\tt rw1(wbc)} and {\tt rw2(tpi)} smoothed predictors with the number of discretized unique values set to 50 for both variables together with the {\tt inla} default base prior G(1, 5e-5) for the precisions of both latent models.
The obtained sensitivity estimates are reported in Table~\ref{TabEx5Sens_and_Sens_smooth_ig_n50nonscaled} (right).
In general, smoothing of {\tt wbc} and {\tt tpi} lead to increased worst-case sensitivity estimates and, for example, the mean change in the unit variance normal distributions for posteriors {\tt rw1(wbc)} in the ``ICAR" increased to 83$\%$ of the mean change in the unit variance normal distributions for priors.
Interestingly, the super-sensitivity for both ``iid" and ``ICAR" models disappeared, although their sensitivity values remained high.
Median sensitivity estimates for precisions of {\tt rw1(wbc)} (0.4) and {\tt rw2(tpi)} (0.86) in ``ICAR" model indicate that although the worst-case sensitivity comparable, the amount of sensitivity in all directions can considerably differ.
We conclude that {\tt rw2(tpi)} is more sensitive at the base prior parameter specification than {\tt rw1(wbc)}.
Further analyses using scaled prior parameter values as suggested by \citet{sorbye_rue_13} can be found in Supplementary Material.

% =======================================
\section{Discussion} \label{S5}
% =======================================

We introduced and utilized a new formal local robustness measure, which was able to automatically handle both circular and worst-case sensitivities in complex Bayesian hierarchical models.
It hinged on two essential ingredients.
First, on an appropriately generated grid, which provided a well defined (standardized) way for modification of the base prior parameter values.
Second, on an instantaneous computation (without any re-run of the model) of the marginal posterior density for priors with parameter values contained in the grid.

It is a formal local robustness approach which dispenses with Taylor-expansion approximation and infinitesimal asymptotics.
Instead, the circular sensitivity is computed for each polar direction chosen equidistantly on $[-\pi, \pi]$ around $\bm{\gamma}_{0}$.
It can be both plotted and summarized by a single number.
Whereas infinitesimal methods restrict their output to the worst-case value, the deeper insight provided by the circular sensitivity and its versatile summaries seems to be new.

We provided a careful and extensive investigation of the properties of the introduced circular sensitivity on conjugate examples, where estimation of exact analytical sensitivity estimates was possible.
We detected a very good agreement of the {\tt inla}-driven and exact sensitivity estimates.
Due to limited space we reported mostly on the worst-case values.
However, for all of the examples polar plots of the circular sensitivity estimates and other summaries could have been accessed immediately for more careful investigations.

As expected, a strong influence of the sample size on the prior sensitivity estimates emerged.
Indeed, we observed that our measure automatically adjusts for increasing sample size by returning smaller prior sensitivity estimates.
The choice of the $\epsilon$ for grid search did not have much influence on sensitivity estimates but anchored calibration in terms of unit-variance normal distribution with shifted means.
What is more, our novel calibration use gave rise to a convenient interpretation of sensitivity estimates independent of the actual $\epsilon$ choice.

We identified several model components and base parameter values specifications requiring more careful attention.
Consideration of miscellaneous complex Bayesian hierarchical models including ``iid", ``ICAR", ``spde", ``rw1" and ``rw2" latent models on several different data sets lead to even more exciting findings.
Sensitivity estimates in Section~\ref{S4.4} indicated clear identifiability problems when both ``iid" and ``ICAR" models were included in the model simultaneously.
We believe that inclusion of both latent models at the same time lead to an overparametrized model.
More super-sensitivities were found in Section~\ref{S4.5} for ``iid" and ``ICAR" latent models.
In addition, smoothing of two covariates with ``rw1" and ``rw2" lead to clearly increased sensitivity estimates.
To our surprise we found no overly elevated sensitivity values for ``spde" components hyperpriors.

Apart of that we developed a handy package {\tt priorSens} for routine, every-day sensitivity computation.
It can be used with practically no extra programming effort needed for default sensitivity investigations within {\tt inla} or to obtain standardized grids for any alternative robustness considerations.
At the moment normal and gamma priors are supported by the package covering already a great amount of possible models.
Further extensions to support other priors are possible.
The {\tt priorSens} package can be obtained upon request from the authors.
We plan to include it as a default option in the {\tt R-INLA} framework.

One possible drawback of our approach is that we investigated local sensitivity for each model component separately while keeping all other model component prior parameters fixed at their base values.
It can happen, however, that a model is insensitive to changes in only one input at time, while being sensitive to simultaneous changes in more than one input.
We believe, however, that local sensitivity for each model component separately is really what we are able to interpret in practice.

Another possible drawback of our approach is that it hinges on the choice of $\epsilon$ for the grid search surrounding the base prior parameter values.
Luckily, we were able to show that sensitivity estimates stay numerically stable over a wide range of $\epsilon$ values leaving much freedom for its choice.
We stress, however, that it is essential to apply the identical $\epsilon$ for all model components for which local sensitivity is examined in order to provide well standardized robustness comparisons.

According to \citet{berger_riosinsua_ruggeri_00} the MCMC methodology is not directly compatible with many of the robust Bayesian techniques that had been developed, so that it is unclear how many of the formal sensitivity measures could be incorporated into the MCMC framework.
\citet{lesaffre_lawson_12} admit that a routine use of sensitivity procedures in MCMC cannot be afforded due to a substantial computational burden.
In contrast, our fast local circular sensitivity estimation technique has a potential to be implemented without much extra cost by any framework capable of estimation of marginal posterior densities.
In particular, it could be computed also with MCMC, where density estimates of $\pi_{\gamma _0} (\theta|y)$ \citep{gelman_carlin_stern_rubin_04} are obtained from MCMC samples.
The usability of our approach could be, therefore, extended far beyond the {\tt R-INLA} applicability range.

In general, our formal local sensitivity measure gave novel, reasonable, easy to interpret and handy piece of information about the marginal posterior distribution sensitivity to base prior parameter values.
Its use was not restricted to conjugate examples but was easily extended to complex Bayesian hierarchical models revealing new insight in identifiability of model components given the data at hand.
Besides, we were able to spot in applications which model components were hard to learn from the data and identified several base prior values specifications requiring more careful attention.
Therefore, we believe that thank to our formal sensitivity measure and the {\tt priorSens} package checking for local robustness in complex Bayesian hierarchical models will become a part of routine statistical practice.

% =======================================
\section{Acknowledgments} \label{Acknowledgments}
% =======================================

We are grateful to Finn Lindgren, Daniel Simpson, Daniel Saban{\'e}s Bov{\'e} and Andrea Riebler for enlightening discussions and to Diego Morosoli (librarian) for acquiring references.

% =======================================
\section{Appendix A (Proof)} \label{Appendix1}
% =======================================

%\noindent{\bf Proof.\ }\qquad
\noindent{\bf Proof of Equation~(\ref{detRW1}):} The determinant of the tridiagonal matrix $\mathbf{Q}=\tau\mathbf{R}+\kappa\mathbf{I}$ with values
\begin{equation*}
\mathbf{Q}_{ij}=\left\{
\begin{array}{rcl}
\tau+\kappa &\ & \mbox{if $i=j=1, n$,} \\
2\tau+\kappa &\ & \mbox{if $1<i=j<n$,} \\
-\tau &\ & \mbox{if $i=j+1, j-1$,} \\
0 &\ & \mbox{otherwise.}
\end{array}
\right.
\end{equation*}
can be computed explicitly by the following argument:

According to \citet[equation (3.23)]{rue_held_05} the eigenvalues of the tridiagonal matrix $\mathbf{R}$ are equal to
\begin{equation*}
\lambda_{i}=2-2\cos(\pi(i-1)/n), i=1,\ldots, n,\ \text{with}\ \lambda_{1}=0.
\end{equation*}
Note that the eigenvalues of the matrix $\tau\mathbf{R}$ are equal to $\lambda_{i}^{*}=\tau\lambda_{i}$.
By \citet[p. 467]{mardia_kent_bibby_79} the eigenvalues of the matrix $\mathbf{Q}=\tau\mathbf{R}+\kappa\mathbf{I}$ are equal to $\lambda_{i}^{**}=\tau\lambda_{i}+\kappa$.
Therefore, the determinant of the tridiagonal matrix $\mathbf{Q}$ reads
\begin{equation*}
|\mathbf{Q}|=|\tau\mathbf{R}+\kappa\mathbf{I}|=\prod_{i=1}^{n}(\tau\lambda_{i}+\kappa)=\prod_{i=1}^{n}(\tau(2-2\cos(\pi(i-1)/n))+\kappa).
\end{equation*}
\qquad $\qedbox$

% =======================================
\section{Appendix B ({\tt R-INLA} review)} \label{AppendixINLA}
% =======================================

A wide range of Bayesian problems is covered by the latent Gaussian models framework and therefore effectively handled by INLA \citep{rue_martino_chopin_09}.
An {\tt R} package (\url{http://www.r-inla.org}) called INLA serves as an interface to the {\tt inla} program.
Its usage is similar to the familiar user-friendly {\tt glm} function in {\tt R}.
The {\tt inla} program allows the user to conveniently perform approximate Bayesian inference in latent Gaussian models.
It is a fast and very versatile program, providing full Bayesian analysis of GLMMs \citep{fong_rue_wakefield_09, martins_simpson_lindgren_rue_13}.
Computationally expensive models on high-dimensional data within stochastic partial differential equations (SPDEs) framework \citep{lindgren_rue_lindstroem_11} can be tackled by {\tt inla} as well.
As output marginal posterior densities of all parameters in the model together with summary characteristics are offered by default.
Although {\tt inla} provides diagnostics for outlying observations via the conditional predictive ordinate ($\cpo$) \citep{pettit_90, geisser_93} default prior sensitivity diagnostics are still missing.
Here, we closed this gap and provided a ready to use {\tt priorSens} package in {\tt R}.

% =======================================
%\section{References} \label{References}
% =======================================

\bibliographystyle{elsarticle-harv}
\bibliography{salbib_20131216}

\begin{thebibliography}{75}
\expandafter\ifx\csname natexlab\endcsname\relax\def\natexlab#1{#1}\fi
\expandafter\ifx\csname url\endcsname\relax
  \def\url#1{\texttt{#1}}\fi
\expandafter\ifx\csname urlprefix\endcsname\relax\def\urlprefix{URL }\fi

\bibitem[{Amari(1990)}]{amari_90}
Amari, S., 1990. Differential-Geometrical Methods in Statistics. 2nd Edition.
  Lecture Notes in Statistics. Vol.~28. Springer-Verlag.

\bibitem[{Amari and Nagaoka(2000)}]{amari_nagaoka_00}
Amari, S., Nagaoka, H., 2000. Methods of {I}nformation {G}eometry. Oxford
  University Press.

\bibitem[{Banerjee et~al.(2004)Banerjee, Carlin, and
  Gelfand}]{banerjee_carlin_gelfand_04}
Banerjee, D., Carlin, B., Gelfand, A., 2004. Hierarchical Modeling and Analysis
  for Spatial Data. Chapman $\&$ Hall.

\bibitem[{Berger et~al.(2000)Berger, {R{\'i}os Insua}, and
  Ruggeri}]{berger_riosinsua_ruggeri_00}
Berger, J.~O., {R{\'i}os Insua}, D., Ruggeri, F., 2000. Bayesian robustness.
  In: {R{\'i}os Insua}, D., Ruggeri, F. (Eds.), Robust Bayesian Analysis.
  Springer-Verlag, pp. 1--32.

\bibitem[{Besag et~al.(1991)Besag, York, and Molli{\'e}}]{besag_york_mollie_91}
Besag, J., York, J., Molli{\'e}, A., 1991. Bayesian image restoration, with two
  applications in spatial statistics. Annals of the Institute of Statistical
  Mathematics 43~(1), 1--59.

\bibitem[{Bhattacharyya(1943)}]{bhattacharyya_43}
Bhattacharyya, A., 1943. On a measure of divergence between two statistical
  populations defined by their probability distributions. Bulletin of the
  Calcutta Mathematical Society 35, 99--109.

\bibitem[{Box(1980)}]{box_80}
Box, G., 1980. Sampling and {B}ayes' inference in scientific modelling and
  robustness. Journal of the Royal Statistical Society, Series A. 143~(4),
  383--430.

\bibitem[{Breslow and Clayton(1993)}]{breslow_clayton_93}
Breslow, N.~E., Clayton, D.~G., 1993. Approximate inference in generalized
  linear mixed models. Journal of the American Statistical Association
  88~(421), 9--25.

\bibitem[{Cacuci(2003)}]{cacuci_03}
Cacuci, D.~G., 2003. Sensitivity and Uncertainty Analysis. {V}olume I,
  {T}heory. Chapman $\&$ Hall.

\bibitem[{Cacuci et~al.(2005)Cacuci, Ionescu-Bujor, and
  Navon}]{cacuci_ionescubujor_navon_05}
Cacuci, D.~G., Ionescu-Bujor, M., Navon, I.~M., 2005. Sensitivity and
  Uncertainty Analysis. {V}olume II, {A}pplications to Large-Scale Systems.
  Chapman $\&$ Hall.

\bibitem[{Cameletti et~al.(2012)Cameletti, Lindgren, Simpson, and
  Rue}]{cameletti_lindgren_simpson_rue_12}
Cameletti, M., Lindgren, F., Simpson, D., Rue, H., 2012. Spatio-temporal
  modeling of particulate matter concentration through the {SPDE} approach.
  AStA Advances in Statistical Analysis, 1--23.
\newline\urlprefix\url{http://dx.doi.org/10.1007/s10182-012-0196-3}

\bibitem[{Carlin and Louis(1998)}]{carlin_louis_98}
Carlin, B., Louis, T., 1998. Bayes and Empirical Bayes Methods for Data
  Analysis. Chapman $\&$ Hall/CRC.

\bibitem[{Clarke and Gustafson(1998)}]{clarke_gustafson_98}
Clarke, B., Gustafson, P., 1998. On the overall sensitivity of the posterior
  distribution to its inputs. Journal of Statistical Planning and Inference
  71~(1-2), 137--150.

\bibitem[{Clayton and Kaldor(1987)}]{clayton_kaldor_87}
Clayton, D., Kaldor, J., 1987. Empirical {B}ayes estimates of age-standardized
  relative risks for use in disease mapping. Biometrics 43~(3), 671--681.

\bibitem[{Cook(1986)}]{cook_86}
Cook, R., 1986. Assessment of local influence. Journal of the Royal Statistical
  Society, Series B. 48~(2), 133--169.

\bibitem[{Dawid(1977)}]{dawid_77}
Dawid, A., 1977. Further comments on some comments on a paper by {B}radley
  {E}fron. The Annals of Statistics 5~(6), 1249.

\bibitem[{Dawid(1979)}]{dawid_79}
Dawid, A., 1979. Conditional independence in statistical theory. Journal of the
  Royal Statistical Society. Series B (Methodological) 41~(1), 1--31.

\bibitem[{Dey and Birmiwal(1994)}]{dey_birmiwal_94}
Dey, D., Birmiwal, L., 1994. Robust {B}ayesian analysis using divergence
  measures. Statistics $\&$ Probability Letters 20~(4), 287--294.

\bibitem[{Eberly and Carlin(2000)}]{eberly_carlin_00}
Eberly, L., Carlin, B., 2000. Identifiability and convergence issues for
  {M}arkov chain {M}onte {C}arlo fitting of spatial models. Statistics in
  Medicine 19~(17-18), 2279--2294.

\bibitem[{Evans and Moshonov(2006)}]{evans_moshonov_06}
Evans, M., Moshonov, H., 2006. Checking for prior-data conflict. Bayesian
  Analysis 4~(1), 893--914.

\bibitem[{Fong et~al.(2010)Fong, Rue, and Wakefield}]{fong_rue_wakefield_09}
Fong, Y., Rue, H., Wakefield, J., 2010. Bayesian inference for generalized
  linear mixed models. Biostatistics {11}~({3}), 397--412.

\bibitem[{Geisser(1992)}]{geisser_92}
Geisser, S., 1992. Bayesian perturbation diagnostics and robustness. In: Goel,
  P., Iyengar, N. (Eds.), Bayesian Analysis in Statistics and Econometrics.
  Springer-Verlag, pp. 289--301.

\bibitem[{Geisser(1993)}]{geisser_93}
Geisser, S., 1993. Predictive {I}nference: {A}n {I}ntroduction. Chapman $\&$
  Hall, Inc.

\bibitem[{Gelfand and Sahu(1999)}]{gelfand_sahu_99}
Gelfand, A., Sahu, S., 1999. Identifiability, improper priors, and {G}ibbs
  sampling for {G}eneralized {L}inear {M}odels. Journal of the American
  Statistical Association 94~(445), 247--253.

\bibitem[{Gelman et~al.(2004)Gelman, Carlin, Stern, and
  Rubin}]{gelman_carlin_stern_rubin_04}
Gelman, A., Carlin, J., Stern, H., Rubin, D., 2004. Bayesian Data Analysis. 2nd
  Edition. Chapman $\&$ Hall/CRC.

\bibitem[{Gilks et~al.(1996)Gilks, Richardson, and
  Spiegelhalter}]{gilks_richardson_spiegelhalter_96}
Gilks, W., Richardson, S., Spiegelhalter, D., 1996. Markov {C}hain {M}onte
  {C}arlo. Chapman $\&$ Hall.

\bibitem[{Goutis and Robert(1998)}]{goutis_robert_98}
Goutis, C., Robert, C., 1998. Model choice in generalised linear models: {A}
  {B}ayesian approach via {K}ullback-{L}eibler projections. Biometrika 85~(1),
  29--37.

\bibitem[{Gustafson(2000)}]{gustafson_00}
Gustafson, P., 2000. Local robustness in {B}ayesian analysis. In: {R{\'i}os
  Insua}, D., Ruggeri, F. (Eds.), Robust Bayesian Analysis. Springer-Verlag,
  pp. 71--88.

\bibitem[{Gustafson and Wasserman(1995)}]{gustafson_wasserman_95}
Gustafson, P., Wasserman, L., 1995. Local sensitivity diagnostics for
  {B}ayesian inference. The Annals of Statistics 23~(6), 2153--2167.

\bibitem[{Harvey(1989)}]{harvey_1989}
Harvey, A., 1989. Forecasting, {S}tructural {T}ime {S}eries {M}odels and the
  {K}alman {F}ilter. Cambridge: Cambridge University Press.

\bibitem[{Harvey and Durbin(1986)}]{harvey_durbin_1986}
Harvey, A., Durbin, J., 1986. The effects of seat belt legislation on {B}ritish
  road causalities: a case study in structural time series modelling. Journal
  of the Royal Statistical Society, Series A 149~(3), 187--227.

\bibitem[{Held and Rue(2010)}]{held_rue_2010}
Held, L., Rue, H., 2010. Conditional and intrinsic autoregressions. In:
  Gelfand, A., Diggle, P., Fuentes, M., Guttorp, P. (Eds.), Handbook of Spatial
  Statistics. Chapman $\&$ Hall/CRC, pp. 201--216.

\bibitem[{Henderson et~al.(2002)Henderson, Shimakura, and
  Gorst}]{henderson_shimakura_gorst_02}
Henderson, R., Shimakura, S., Gorst, D., 2002. Modeling spatial variation in
  leukemia survival data. Journal of the American Statistical Association
  97~(460), 965--972.

\bibitem[{Ibrahim et~al.(2011)Ibrahim, Zhu, and Tang}]{ibrahim_zhu_tang_11}
Ibrahim, J., Zhu, H., Tang, N., 2011. Bayesian local influence for survival
  models. Lifetime Data Analysis 17~(1), 43--70.

\bibitem[{Jeffreys(1961)}]{jeffreys_61}
Jeffreys, H., 1961. Theory of {P}robability. Oxford University Press.

\bibitem[{Kadane(1992)}]{kadane_for_geisser_92}
Kadane, J., 1992. Comments to: ``{B}ayesian perturbation diagnostics and
  robustness" by {S}. {G}eisser. In: Goel, P., Iyengar, N. (Eds.), Bayesian
  Analysis in Statistics and Econometrics. Springer-Verlag, pp. 298--300.

\bibitem[{Kass et~al.(1989)Kass, Tierney, and Kadane}]{kass_tierney_kadane_89}
Kass, R., Tierney, L., Kadane, J., 1989. Approximate methods for assessing
  influence and sensitivity in {B}ayesian analysis. Biometrika 76~(4),
  663--674.

\bibitem[{Kneib and Fahrmeir(2007)}]{kneib_fahrmeir_07}
Kneib, T., Fahrmeir, L., 2007. A mixed model approach for geoadditive hazard
  regression. Scandinavian Journal of Statistics 34~(1), 207--228.

\bibitem[{Lavine(1992)}]{lavine_92}
Lavine, M., 1992. Local predictive influence in {B}ayesian linear models with
  conjugate priors. Communications in Statistics - Simulation and Computation
  21~(1), 269--283.

\bibitem[{Le~Cam(1986)}]{LeCam_86}
Le~Cam, L., 1986. Asymptotic Methods in Statistical Decision Theory.
  Springer-Verlag.

\bibitem[{Lesaffre and Lawson(2012)}]{lesaffre_lawson_12}
Lesaffre, E., Lawson, A., 2012. Bayesian {B}iostatistics. John Wiley $\&$ Sons.

\bibitem[{Lindgren(2012)}]{lindgren_13}
Lindgren, F., 2012. Continuous domain spatial models in {R}-{INLA}. The ISBA
  Bulletin 19~(4).

\bibitem[{Lindgren et~al.(2011)Lindgren, Rue, and
  Lindstr{\"o}m}]{lindgren_rue_lindstroem_11}
Lindgren, F., Rue, H., Lindstr{\"o}m, J., 2011. An explicit link between
  {G}aussian fields and {G}aussian {M}arkov random fields: the stochastic
  differential equation approach. Journal of the Royal Statistical Society,
  Series B. 73~(4), 423--498.

\bibitem[{Mardia et~al.(1979)Mardia, Kent, and Bibby}]{mardia_kent_bibby_79}
Mardia, K.~V., Kent, J.~T., Bibby, J.~M., 1979. Multivariate Analysis. Academic
  Press.

\bibitem[{Martino et~al.(2011)Martino, Akerkar, and
  Rue}]{martino_akerkar_rue_11}
Martino, S., Akerkar, R., Rue, H., 2011. Approximate {B}ayesian inference for
  survival models. Scandinavian Journal of Statistics 38~(3), 514--528.

\bibitem[{Martins et~al.(2013)Martins, Simpson, Lindgren, and
  Rue}]{martins_simpson_lindgren_rue_13}
Martins, T., Simpson, D., Lindgren, F., Rue, H., 2013. Bayesian computing with
  {INLA}: new features. Computational Statistics \& Data Analysis 67, 68--83.

\bibitem[{McCulloch(1989)}]{mcculloch_89}
McCulloch, R., 1989. Local model influence. Journal of the American Statistical
  Association 84~(406), 473--478.

\bibitem[{Millar and Stewart(2007)}]{millar_stewart_07}
Millar, R., Stewart, W., 2007. Assessment of locally influential observations
  in {B}ayesian models. Bayesian Analysis 2~(2), 365--384.

\bibitem[{M{\"u}ller(2012)}]{mueller_12}
M{\"u}ller, U., 2012. Measuring prior sensitivity and prior informativeness in
  large {B}ayesian models. Journal of Monetary Economics 59~(6), 581--597.

\bibitem[{Oakley and O'Hagan(2004)}]{oakley_ohagan_04}
Oakley, J., O'Hagan, A., 2004. Probabilistic sensitivity analysis of complex
  models: a {B}ayesian approach. Journal of the Royal Statistical Society,
  Series B. 66~(3), 751--769.

\bibitem[{P{\'e}rez et~al.(2006)P{\'e}rez, Mart{\'i}n, and
  Rufo}]{perez_martin_rufo_06}
P{\'e}rez, C., Mart{\'i}n, J., Rufo, M., 2006. {MCMC}-based local parametric
  sensitivity estimation. Computational Statistics $\&$ Data Analysis 51~(2),
  823--835.

\bibitem[{Pettit(1990)}]{pettit_90}
Pettit, L., 1990. The conditional predictive ordinate for the normal
  distribution. Journal of the Royal Statistical Society, Series B. 52~(1),
  175--184.

\bibitem[{Plummer(2001)}]{plummer_01}
Plummer, M., 2001. Local sensitivity in {B}ayesian graphical models.
\newline\urlprefix\url{http://www-ice.iarc.fr/~martyn/papers/sensitivity.ps}

\bibitem[{Rao(1945)}]{rao_45}
Rao, C., 1945. Information and the accuracy attainable in the estimation of
  statistical parameters. Bulletin of the Calcutta Mathematical Society 37,
  81--91.

\bibitem[{{R{\'i}os Insua} et~al.(2000){R{\'i}os Insua}, Ruggeri, and
  Mart{\'i}n}]{riosinsua_ruggeri_martin_00}
{R{\'i}os Insua}, D., Ruggeri, F., Mart{\'i}n, J., 2000. Bayesian sensitivity
  analysis. In: Saltelli, A., Chan, K., Scott, E.~M. (Eds.), Sensitivity
  Analysis. John Wiley $\&$ Sons, pp. 225--244.

\bibitem[{Robert(1996)}]{robert_96}
Robert, C., 1996. Intrinsic losses. Theory and Decision 40~(2), 191--214.

\bibitem[{Roos and Held(2011)}]{roos_held_11}
Roos, M., Held, L., 2011. Sensitivity analysis in {B}ayesian generalized linear
  mixed models for binary data. Bayesian Analysis 6~(2), 259--278.

\bibitem[{Rue and Held(2005)}]{rue_held_05}
Rue, H., Held, L., 2005. Gaussian {M}arkov {R}andom {F}ields. Theory and
  Applications. Chapman $\&$ Hall/CRC.

\bibitem[{Rue et~al.(2009)Rue, Martino, and Chopin}]{rue_martino_chopin_09}
Rue, H., Martino, S., Chopin, N., 2009. Approximate {B}ayesian inference for
  latent {G}aussian models by using integrated nested {L}aplace approximations.
  Journal of the Royal Statistical Society, Series B. 71~(2), 319--392.

\bibitem[{Ruggeri(2008)}]{ruggeri_08}
Ruggeri, F., 2008. Bayesian robustness. Forum: Robustness Analysis, In:
  European Working Group ``Multiple Criteria Decision Aiding" 3~(17).

\bibitem[{Saltelli et~al.(2000)Saltelli, Chan, Scott, and
  (Eds.)}]{saltelli_chan_scott_00}
Saltelli, A., Chan, K., Scott, E.~M., (Eds.), 2000. Sensitivity {A}nalysis.
  John Wiley $\&$ Sons.

\bibitem[{Saltelli et~al.(2008)Saltelli, Ratto, Andres, Campolongo, Cariboni,
  Gatelli, Saisana, and
  Tarantola}]{saltelli_ratto_andres_campolongo_cariboni_gatelli_saisana_tarantola_08}
Saltelli, A., Ratto, M., Andres, T., Campolongo, F., Cariboni, J., Gatelli, D.,
  Saisana, M., Tarantola, S., 2008. Global {S}ensitivity {A}nalysis. {T}he
  {P}rimer. John Wiley $\&$ Sons.

\bibitem[{Saltelli et~al.(2004)Saltelli, Tarantola, Campolongo, and
  Ratto}]{saltelli_tarantola_campolongo_ratto_04}
Saltelli, A., Tarantola, S., Campolongo, F., Ratto, M., 2004. Sensitivity
  {A}nalysis in {P}ractice. {A} {G}uide to {A}ssessing {S}cientific {M}odels.
  John Wiley $\&$ Sons.

\bibitem[{Simpson et~al.(2012{\natexlab{a}})Simpson, Lindgren, and
  Rue}]{simpson_lindgren_rue_12a}
Simpson, D., Lindgren, F., Rue, H., 2012{\natexlab{a}}. In order to make
  spatial statistics computationally feasible, we need to forget about the
  covariance function. Environmetrics 23~(1), 65--74.

\bibitem[{Simpson et~al.(2012{\natexlab{b}})Simpson, Lindgren, and
  Rue}]{simpson_lindgren_rue_12b}
Simpson, D., Lindgren, F., Rue, H., 2012{\natexlab{b}}. Think continuous:
  {M}arkovian {G}aussian models in spatial statistics. Spatial Statistics 1,
  16--29.

\bibitem[{Sivaganesan(2000)}]{sivaganesan_00}
Sivaganesan, S., 2000. Global and local robustness approaches: uses and
  limitations. In: {R{\'i}os Insua}, D., Ruggeri, F. (Eds.), Robust Bayesian
  Analysis. Springer-Verlag, pp. 89--108.

\bibitem[{S{\o}rbye and Rue(2013)}]{sorbye_rue_13}
S{\o}rbye, S., Rue, H., 2013. Scaling intrinsic {G}aussian {M}arkov random
  field priors in spatial modelling. Spatial Statistics.

\bibitem[{Tierney and Kadane(1986)}]{tierney_kadane_86}
Tierney, L., Kadane, J., 1986. Accurate approximations for posterior moments
  and marginal densities. Journal of the American Statistical Association
  81~(393), 82--86.

\bibitem[{Tierney et~al.(1989)Tierney, Kass, and
  Kadane}]{tierney_kass_kadane_89}
Tierney, L., Kass, R., Kadane, J., 1989. Fully exponential {L}aplace
  approximations to expectations and variances of nonpositive functions.
  Journal of the American Statistical Association 84~(407), 710--716.

\bibitem[{{Van der Linde}(2007)}]{vanderLinde_07}
{Van der Linde}, A., 2007. Local influence on posterior distributions under
  multiplicative modes of perturbation. Bayesian Analysis 2~(2), 319--332.

\bibitem[{Wakefield(2007)}]{wakefield_07}
Wakefield, J., 2007. Disease mapping and spatial regression with count data.
  Biostatistics 8~(2), 158--183.

\bibitem[{Weiss(1996)}]{weiss_96}
Weiss, R., 1996. An approach to {B}ayesian sensitivity analysis. Journal of the
  Royal Statistical Society, Series B. 58~(4), 739--750.

\bibitem[{Weiss and Cook(1992)}]{weiss_cook_92}
Weiss, R., Cook, R., 1992. A graphical case statistics for assessing posterior
  influence. Biometrika 79~(1), 51--55.

\bibitem[{Zhu et~al.(2007)Zhu, Ibrahim, Lee, and
  Zhang}]{zhu_ibrahim_lee_zhang_07}
Zhu, H., Ibrahim, J., Lee, S., Zhang, H., 2007. Perturbation selection and
  influence measures in local influence analysis. The Annals of Statistics
  35~(6), 2565--2588.

\bibitem[{Zhu et~al.(2011)Zhu, Ibrahim, and Tang}]{zhu_ibrahim_tang_11}
Zhu, H., Ibrahim, J., Tang, N., 2011. Bayesian influence analysis: a geometric
  approach. Biometrika 98~(2), 307--323.

\end{thebibliography}

\end{document}